\documentclass{elsart}
\usepackage{graphicx}% Include figure files
\usepackage{bm}% bold math
\usepackage{array}
\usepackage{amssymb}

\def \text{\mbox}
\def \p{ \partial}

\def\grapprox{\mathop{\lower.5ex \hbox{$\buildrel{\fivesy >}\over{\fivesy\sim}$}} \nolimits}
\def\lsapprox{\mathop{\lower.5ex \hbox{$\buildrel{\fivesy <}\over{\fivesy\sim}$}} \nolimits}
\def\grls{\mathop{\lower.5ex \hbox{$\buildrel{\fivesy >}\over{\fivesy <}$}} \nolimits}

\def\vec#1{{\bf #1}}

\def\abs#1{\left\vert #1 \right\vert}

\def\pt{\partial}

\def\pxx#1{{\partial #1\over\partial x}}
\def\pyy#1{{\partial #1\over\partial y}}

\def\pss#1{{\partial #1\over\partial s}}

\def\ptt#1{{\partial #1\over\partial t}}

\def\dtt#1{{d #1\over dt}}

\def\grad{\nabla}

\def\dpl{\grad_\parallel}

\def\vexb{\vec v_E}

\def\vedl{\vexb\cdot\grad}

\def\vor{\grad_\perp^2\phi}

\def\bhat{\hat\beta}
\def\muhat{\hat\mu}
\def\epss{\hat\epsilon}

\def\vor{\omega}
\def\Psi{A_\parallel}
\begin{document}

\begin{frontmatter}


\title{ Statistical properties of transport in plasma  turbulence}
\author{Volker Naulin},
\ead{volker.naulin@risoe.dk}
\author{Odd Erik Garcia},
\author{Anders Henry Nielsen},
\author{Jens Juul Rasmussen}
\address{Association EURATOM -- Ris{\o} National Laboratory,
 Optics and Fluid Dynamics Department, OFD-128 Ris{\o},
4000 Roskilde, 
Denmark}

\date{\today}% It is always \today, today,

\begin{abstract}
The statistical properties of the $E\times B$ flux in different types of
plasma turbulence simulations  
are investigated using probability density distribution
functions (PDF). 
The physics included  in the models ranges from two dimensional drift-wave
turbulence to three dimensional MHD simulations.  
The PDFs of the flux surface averaged transport are in good agreement with
an extreme value distribution (EVD). This is interpreted as a signature of
localized vortical structures dominating the transport on a flux surface.  
Models
accounting  for a local input of energy and including the global evolution of 
background profiles show a more bursty behaviour  in 2D as well
as 3D simulations of  interchange and drift  type  turbulence.
Although the PDFs are not Gaussian, again no  
algebraic tails are observed. The tails are, in all cases considered, in excellent
agreement with EVDs.
\end{abstract}

%\pacs{ 52.35.Py, 52.25.Fi, 52.35.Ra, 52.65.-y}

\end{frontmatter}

\section{Introduction} 
In hot magnetised plasmas the main cross-field
transport is often found to be  far larger than expected for diffusive
transport due to collisions. 
In plasma physics the transport is then  called anomalous
while  in fluid dynamics "strange" or "anomalous" transport signifies
the non-diffusive character of the transport.
We investigate the statistical properties of the turbulent particle
flux with the aim to examine
the anomality of the transport in the fluid sense of
meaning.  
An important feature stressed by many authors is the 
non-Gaussianty of the PDF of the point-wise measured flux 
\cite{Huld:Iizuka:Pecseli:Rasmussen:1988,Huld:Nielsen:Pecseli:Rasmussen:1991,Endler:Niedermeyer:etal:1995,Carreras:etal:1996}.
While the non-Gaussianity of the flux PDF is to be expected from a
quantity that itself is a folding of two statistical variables, namely
density and radial velocity, 
the observation of power law tails in the flux PDF would be an
indication of anomalous transport in the fluid sense of the wording, that is non-diffusive transport
behaviour. Correspondingly this would indicate the presence of  long range correlations in the
turbulence \cite{Zaslavsky:2002}. 
Experimental results showing finite size scaling and
similarity of transport PDFs measured in different experimental  devices have recently been
reported in Refs.~\cite{Carreras:Lynch:LaBombard:2001,Hidalgo:2002}. 
These measurements have shown that transport in the edge and scrape-off
layer (SOL) region of fusion devices is highly intermittent, self
similar and does in its statistical features not vary much between
machines of different sizes and geometries.
A  characteristic feature of the transport seems to be the
non-Gaussianity observed in its PDF and its intermittency.
Intermittent transport would indicate that indeed the particle
transport is anomalous  not only in the sense that it is not due to
collisions, but  also in the sense that it is non-diffusive. A
non-diffusive transport, however, has distinct scaling properties,  which
would pose much stricter conditions on the design of the vessel surrounding
a fusion device, as this has  to withstand the largest likely
transport event. 
Additionally, a number of fundamental questions in relation to
transport PDFs are  still 
unanswered, so it is as yet  unclear what ingredients are
needed to produce a certain transport PDF. Are transport PDFs
dependent on the driving mechanism behind the turbulence? How does
different physics enter into the transport PDF? Here we present
initial answers 
to some of these questions by analysing direct numerical
simulations of various types of plasma turbulence in different
geometries. 
%The paper is organized as follows: In Section 2 we consider models
%based on turbulent fluctuations in 2D and 3D, spanning from drift wave
%dynamics in 2D slab geometry to 3D electromagnetic drift-Alfv{\'e}n
%turbulence in flux tube geometry  in the drift and the
%MHD-limits. 
%Low frequency electrostatic fluctuations of this type 
%are the most likely candidate for understanding and modelling  plasma
%transport (see, for example 
%\cite{dw:wagner93})  at the plasma edge, while electromagnetic
%effects can become  important 
%for the instability mechanism driving the turbulence in the edge
%region at relatively low plasma beta \cite{Scott:1997:2}.
% In
%Section 3 we then turn to global models integrating the evolution of
%fluctuations and background and consider a relatively simple 2D
%interchange turbulence model and a 3D drift wave turbulence system in
%terms of the associated transport. Finally we present conclusions in
%Section 4. 

\section{Local turbulence Models}
In this section we consider plasma turbulence in a classical sense, namely
as being based on a scale separation between the equilibrium and
fluctuations,  f.x.~density fluctuations are assumed small
compared the the equilibrium density $\tilde n/n_{00} << 1 $.
The fluid equations for drift  micro-turbulence 
result from standard ordering based upon
the slowness of the dynamics compared to the ion gyro-frequency
$\Omega_i=eB/M_i$ and hence
the smallness of the drift scale $\rho_s$ compared to the background
pressure gradient scale length $L_\perp$.  These quantities and the sound speed $c_s$ are
defined by
\begin{equation}
  \Omega_i = {eB\over M_i }, \qquad 
  c_s^2 = {T_e\over M_i}, \qquad  
  \rho_s = {c_s\over\Omega_i}, \qquad
  L_\perp = \abs{\nabla\log p_e}^{-1},
\end{equation}
where subscripts $e$ and $i$ refer to electrons and ions respectively, 
and the temperature
is given in units of energy.  Normalization is in terms of scaled
dependent variables (electrostatic potential $e\phi/T_e$, electron
density  $n/n_{00}$, parallel
ion velocity $u/c_s$, parallel electric current $J/n_{00} e c_s$; 
respectively).
In addition the dependent quantities are scaled with 
the small parameter $\delta=\rho_s/L_\perp$, so that we calculate with
quantities of order one.  
The quantity $n_{00}$ is a normalizing density, while $n_0(x)$ 
is the equilibrium plasma density having a finite gradient.
In normalized units the radial profile of the density is 
$\partial_r \log n_0(x) = -1 $. Vorticity is defined via $\vor= \nabla_{\perp}^{2} \phi$.
  
\subsection{Two dimensional drift turbulence}
\begin{figure}[h]
\begin{minipage}[t]{.475\textwidth}
\centering
\includegraphics[width=\textwidth]{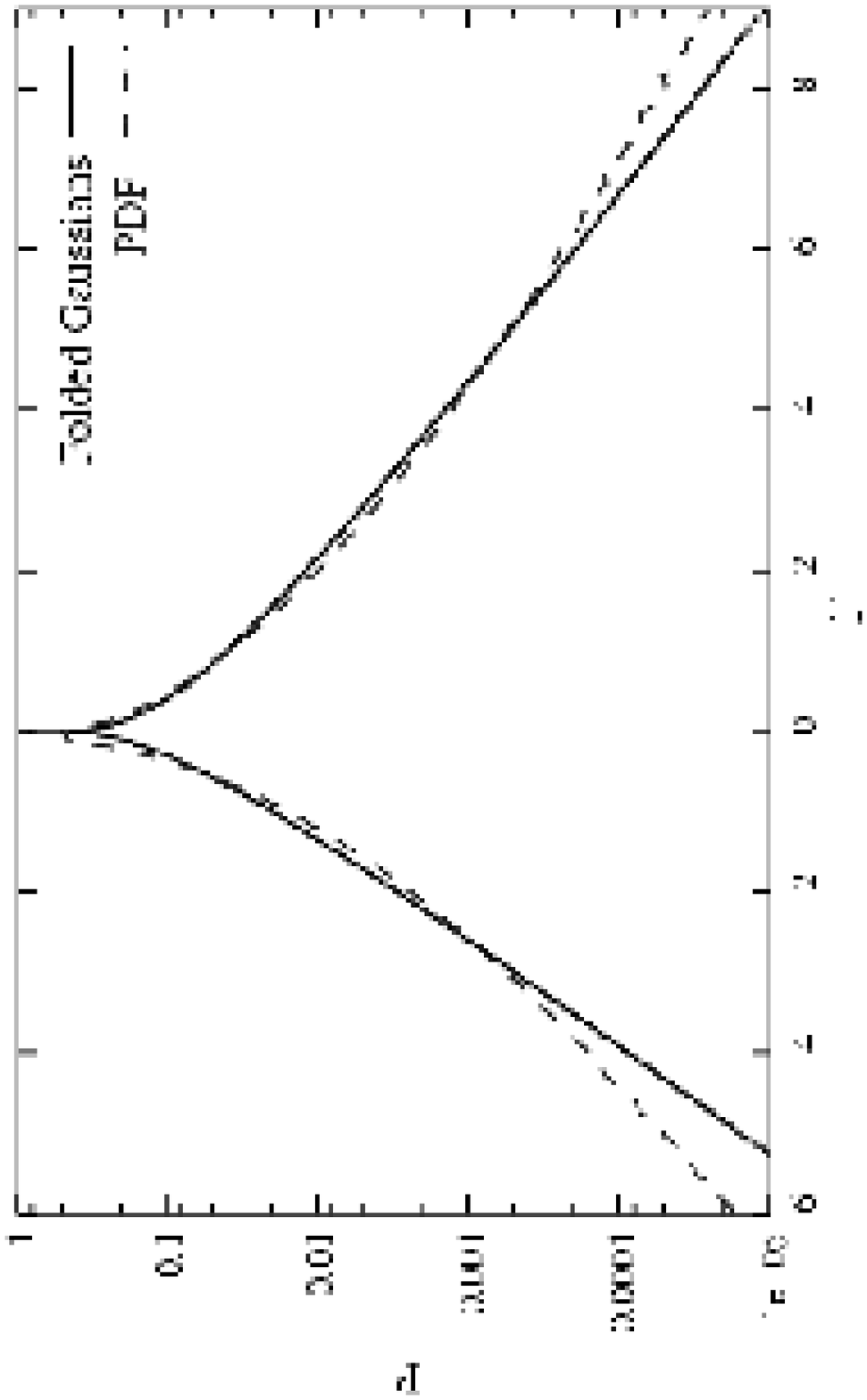}
\caption{PDF of $\Gamma_{n,\text{loc}}$ compared with folded Gaussian}
\label{Fig.pdf1}
\end{minipage}
\hfill
\begin{minipage}[t]{.475\textwidth}
\centering
 \includegraphics[width=\textwidth]{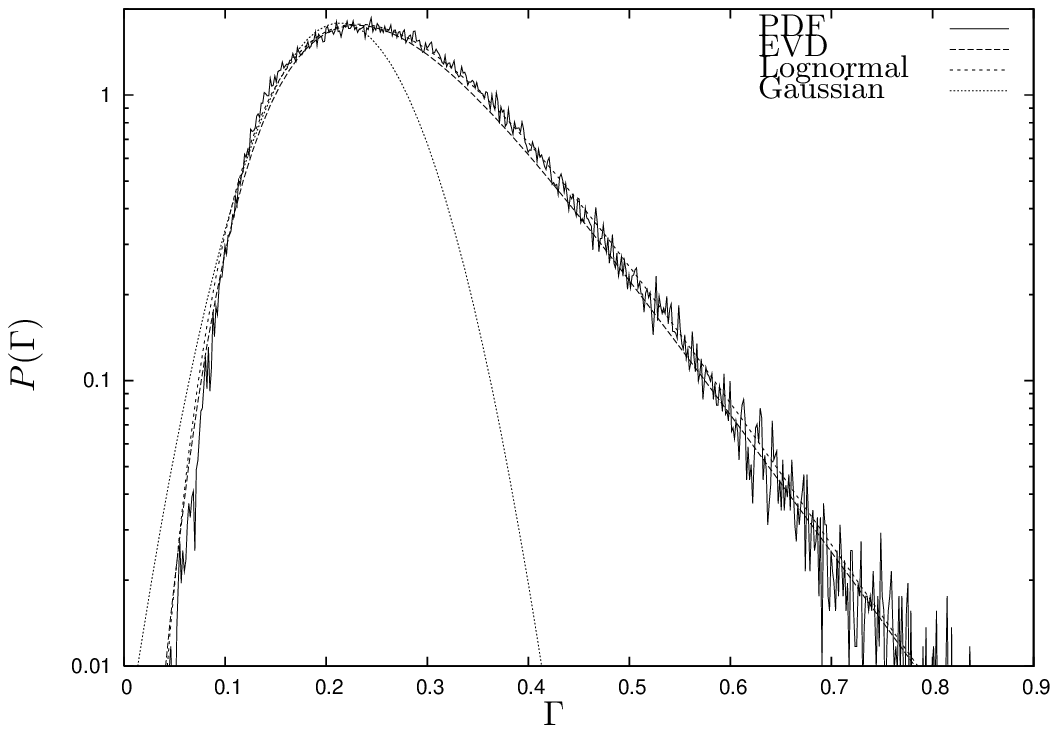}
\caption{PDF of $\Gamma_{n,\text{FS}}$, compared to  an EVD a log-normal
  distribution and a Gaussian} \label{Fig:pdf2}
\end{minipage}
\end{figure}

Assuming electrostatic motion, neglecting eletron inertia and
replacing parallel derivatives by an effective parallel wavelength
$L_{\shortparallel} $ we
arrive at the well known Hasegawa-Wakatani equations
\cite{Wakatani:Hasegawa:1984} (HWE), 
describing two
dimensional drift-wave turbulence in the absence of magnetic field
line curvature:
\begin{equation}
\partial_{t} \vor + \{ \varphi, \vor \}
   =  - \frac{1}{L_\|^2 \nu}  \left(n-\varphi\right)  + \mu_\vor
   \nabla^2 \vor,
\label{EQ:HWOM}
\end{equation}
\begin{equation}
\partial_{t} n + \{ \varphi, n \} + \partial_{y} \varphi   =
- \frac{1}{L_\|^2 \nu}  \left(n-\varphi \right) + \mu_n
   \nabla^2 n\; ,
\label{EQ:HWD}
\end{equation}
where the Poisson bracket
$
{\left\{ f,g \right\} } = \pxx{f}\pyy{g}-\pyy{f}\pxx{g}
$
is used to write the non-linear terms originating from advection with
the $E \times B$ drift velocity.
In this system the turbulence is driven by the resistive
instability. We solve  the HWE  on
a double periodic domain \cite{Naulin:Nielsen:Rasmussen:1999} and typically we
use $256 \times
256$ modes on a square of side length $L=40 $. To produce 
self-consistent stationary turbulence we start the
simulations from low amplitude ($10^{-6}$) random density
fluctuations and initialise all other fields to zero. 
The fluctuations grow due to the resistive instability and
saturate with amplitudes of order one.
After saturation \cite{Naulin:2002}
we measure the local turbulent radial particle transport
($x$-direction) as given in a
single point: 
\begin{equation}
\Gamma_{n,\text{loc}} =  n v_r\;, 
\end{equation}
and the  flux surface
averaged  flux $\Gamma_{n,\text{FS}}$:
\begin{equation}
\Gamma_{n,\text{FS}} = \frac{1}{L_y} \int_0^{L_y} \Gamma_{n,\text{loc}}\; dy\;,
\end{equation}
which in the two dimensional context is trivially obtained by averaging
over the poloidal coordinate $y$. 
Density and potential fluctuations have PDFs that
are well described by  Gaussians. Consequently  
the PDF of
the $E\times B$ flux, $\Gamma_{n,\text{loc}}$, as depicted in
Fig.~\ref{Fig.pdf1}, is close to the functional form  expected
from the folding of Gaussian PDFs for the density fluctuations
$n$ and the radial $E\times B$-velocity, $v_r = -\partial_y \phi$
\cite{Carreras:etal:1996}:
\begin{equation}
P_{\mbox{FG}}(\Gamma) = \frac{1}{\pi } \frac{\sqrt{1-c^2}}{\sigma_n \sigma_{v_r}} K_0
\left(\frac{|\Gamma|}{\sigma_n \sigma_{v_r}} \right)\exp \left( -c
\frac{\Gamma}{\sigma_n \sigma_{v_r}} \right)\;.
\label{flux_G}
\end{equation}
where $c $ is the correlation between density and radial velocity fluctuations
$
c = - \frac{\langle v_r n \rangle}{\langle v_r^2 \rangle^{1/2} \langle n^2
\rangle^{1/2}} = \cos{\alpha_{n v_r}}
$, $K_0$ is the modified Bessel function of second kind, 
and $\alpha_{nu} $ is the phase angle between $n$ and $v_r$. 
Furthermore we have
$ 
\sigma_{v_r} \equiv  \langle v_r^2 \rangle^{1/2} ( 1- c^2)^{1/2}\;,
$
and correspondingly for $\sigma_n$. 
%-----------------
Using  Eq.(\ref{EQ:HWD})  for an  approximate relationship between density and
potential and the drift-wave dispersion relation to
lowest orderin the long-wavelength
limit  $\omega = \frac{k_y}{1+k_\perp^2}$ , we find:
\begin{equation}
\Gamma_{n,\mbox{\small FS}} = - \frac{1}{L_y} \int_0^{L_y} n \p_y \phi \; dy \approx \frac{L_\|^2 \nu}{L_y}  \int_0^{L_y} \left(
  \p_y \nabla_{\perp} \phi \right)^2 \; dy \ge  0 \,.
\end{equation}
Thus the flux surfaced averaged transport is to lowest order positive
definite, as confirmed by the numerical simulations. 
It is thus the PDF of the  logarithm of the transport, $\ln
\Gamma_{n,\mbox{\small FS}} $,  that 
converges towards a Gaussian, if the central limit theorem applies. 
As seen from Fig.~\ref{Fig:pdf2}  the transport PDF of
$\Gamma_{n,\mbox{\small FS}} $
is indeed very well described by a log-normal
distribution with in the present case zero offset $z= 0 $:
\begin{equation}
P_{\mbox{\small ln}}(\Gamma) = \frac{1}{\sqrt{2 \pi} \sigma} \frac{1}{\Gamma - z}  \exp\left(-\frac{1}{2}
\left( \frac{\ln (\frac{\Gamma -  z}{m})}{\sigma}\right)^2\right)\;.
\end{equation}
%This theorem is important because it tells us that the limiting
%distribution of extreme returns always has the same
%form\x{2014}whatever the distribution of the parent variable from
%which the returns are drawn. It is related to the better-known
%central limit theorem, but applies to the extremes of observations
%rather than their means. 
%
An alternative description of the statistical properties of the flux
can be made in terms of the extreme value distribution (EVD).
For a statistical variable --  here the averaged flux -- 
that is dominated by the minimum or maximum of a large number of random
observations -- here the local flux events --  the EVD arises\cite{Book:Sornette:2000}:
 \begin{equation}
P_{\mbox{\small EVD}}(\Gamma) = \frac{1}{g} \exp\left(-
 \frac{\Gamma -  h_0}{g}\right) 
\exp \left[ -\exp \left(-\frac{\Gamma -  h_0}{g}\right) \right]\,.
\end{equation}
The EVD has recently been used successfully to model and explain statistical
properties of electron pressure fluctuations  in a non-confinement plasma experiment
\cite{Rypdal:Ratynskaia:2003}. As seen from Fig.~\ref{Fig:pdf2} the EVD
explains the observed transport PDF in this case as well as the
lognormal distribution and is indistinguishable from the former. 
One should, however,  note that the EVD decays faster than
the lognormal one and, thus,  is better suited to describe statistical
variables that are limited not only from below, but also from above.
Clearly the maximum transport event in all real world systems is limited by
the system size. 
We interpret  the present situation
in the sense that the flux surface averaged transport is
dominated by extreme events, probably mediated by transport
events linked to vortical structures. 
%In this context it is worthwhile
%to remark that the simulations take place in a setup where the size of
%the simulation box is much larger than the correlation length and that
%the aspect ratio $L_y/L_x$ is large.
The flux PDF itself in this case is naturally skewed, but   
the radial particle transport  is by no means 'strange' in the sense
of not being diffusive on  large time and space scales. 
No power law tail is  observed in the
data.  That indeed the transport is diffusive  is augmented by the fact, that the radial
dispersion of ideal test-particles in 2D Hasegawa-Wakatani turbulence is found
to be asymptotically diffusive as well \cite{Basu:Jessen:Naulin:Rasmussen:2003,Basu:Naulin:Rasmussen:2003}.

\subsection{Drift-Alfv{\'e}n and MHD turbulence}

\begin{figure}[h!]
\begin{minipage}[t]{.475\textwidth}
\centering
\includegraphics[width=\textwidth]{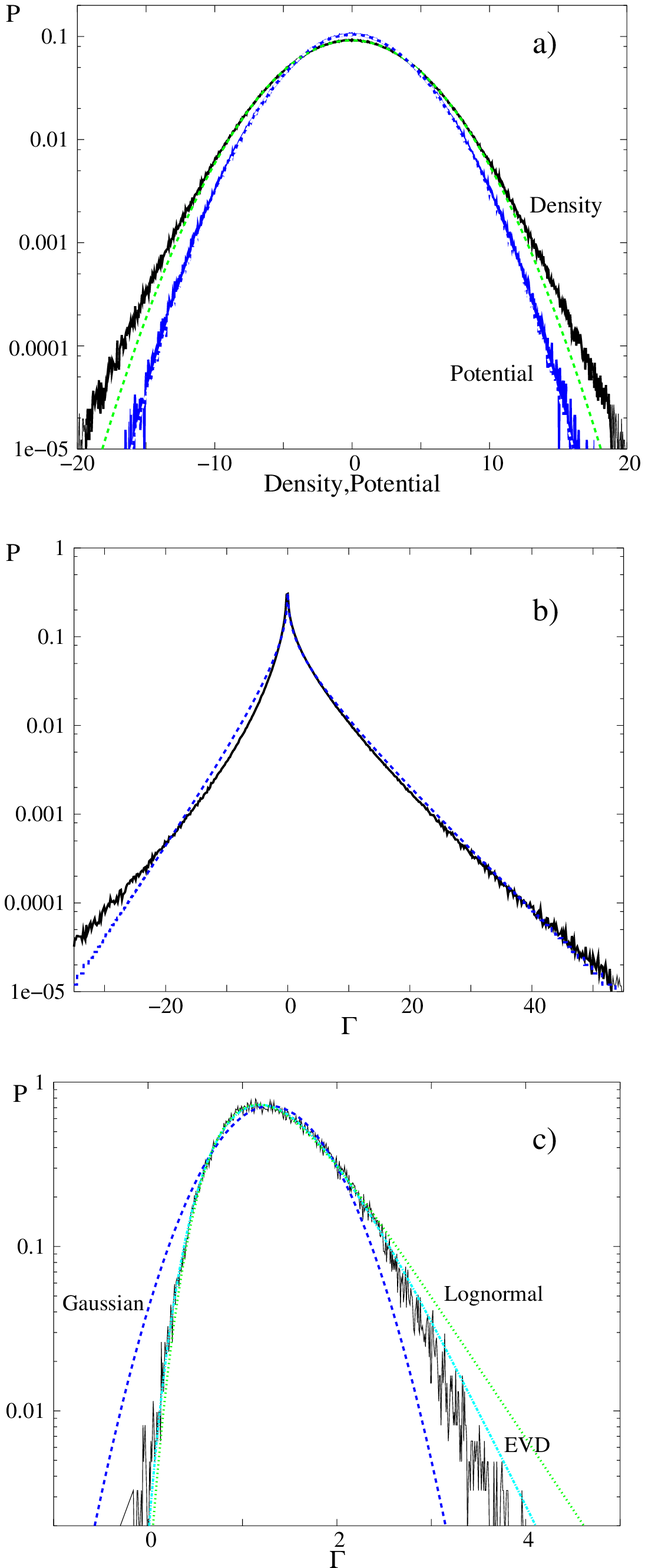}
\caption{
Low $\hat \beta = 0.5$ case. 
a) PDF's for density $n$ and potential $\phi$ fluctuations with
  fitted Gaussians (dashed) 
b) $\Gamma_{n,\text{loc}}$ with fitted folded Gaussian 
c)   $\Gamma_{n,\text{FS}}$ with a fitted Gaussian, log-normal and extreme value
distribution (EVD). 
\label{Fig:FluxPDF_lowbeta} 
}
\end{minipage}
\hfill
\begin{minipage}[t]{.475\textwidth}
\centering
\includegraphics[width=\textwidth]{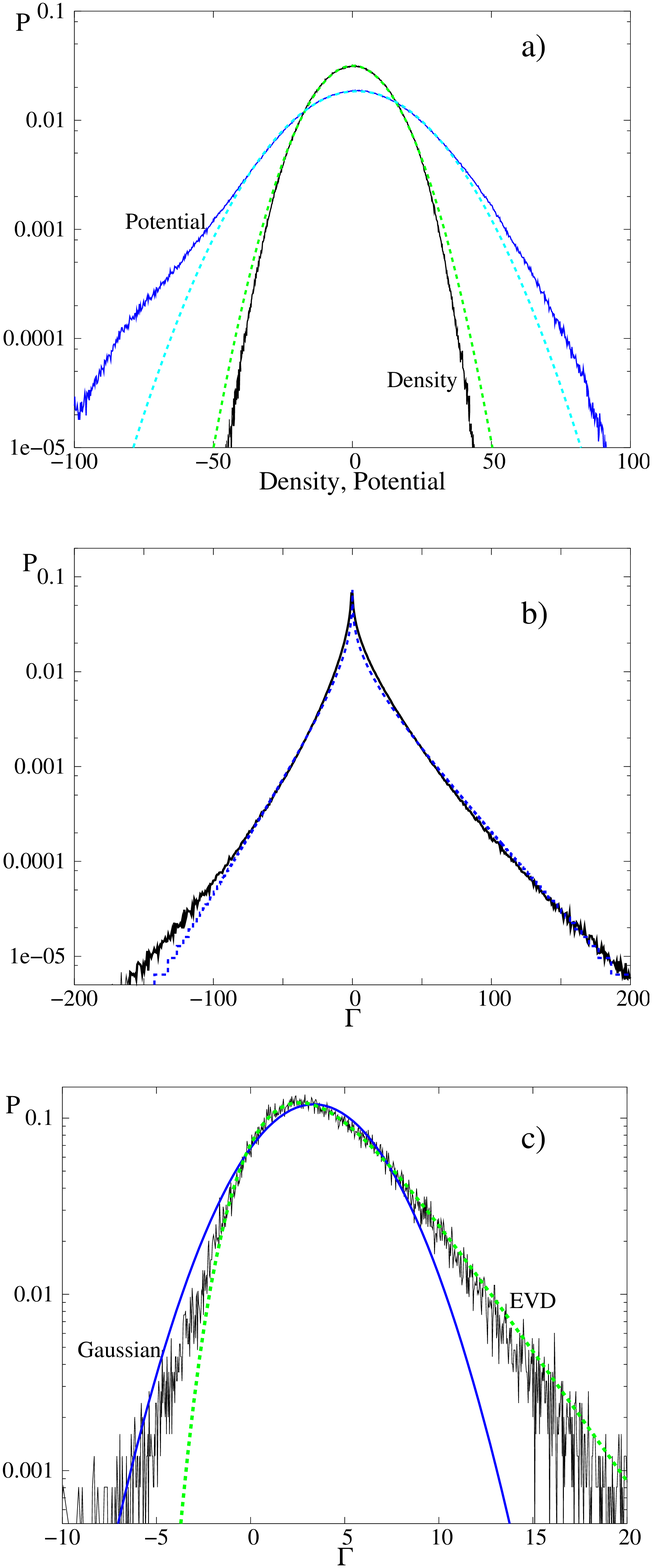}
\caption{High $\hat \beta = 50$ case.
a) PDF's for density $n$ and potential $\phi$ fluctuations with
  fitted Gaussians (dashed) 
b) $\Gamma_{n,\text{loc}}$ with fitted folded Gaussian and
c) $\Gamma_{n,\text{FS}}$ with a fitted Gaussian and extreme value
distribution (EVD). The log-normal distribution is not shown, as it is 
very close to the EVD.
\label{Fig:FluxPDF_highbeta} 
}
\end{minipage}
\end{figure}

Next we  consider electromagnetic drift-Alfv{\'e}n turbulence with
magnetic field curvature effects.
The geometry is a three dimensional flux tube with local slab-like
coordinates $\left[x,y,s\right]$ 
\cite{Scott:2001} and
the model is defined by the 
temporal evolution of the following four
scalar
fields: electrostatic potential $\phi$, 
density $n$ given by electron density continuity equation, 
parallel current $J$, and parallel
ion velocity $u$,
with the 
parallel magnetic potential $\Psi =  - \nabla_{\perp}^{-2} J$:
\begin{equation}
\label{eq:eqvor}
\ptt \vor + \vedl \vor 
=
{ K } \left(  n \right)
+  \nabla_{\|} J  + \mu_{w} \nabla_{\perp}^{2}  \vor\,,
\end{equation}
\begin{equation}
\label{eq:eqne}
\ptt n + \vedl (n_0 + n)
= 
{K} \left( n  - \phi \right)
+  \nabla_{\|} \left( J  -  u\right) +  \mu_{n}  \nabla_{\perp}^{2}  n\,,
\end{equation}
\begin{equation}
\label{eq:eqpsi}
\ptt{} \left( \bhat\Psi + \muhat J \right) + \muhat\vedl J
=
\nabla_{\|} \left( n_0 + n - \phi \right) - C J\,,
\end{equation}
\begin{equation}
\label{eq:equi}
\epss\left(\ptt u + \vedl u\right)
=
 - \nabla_{\|} \left(n_0 + n\right) \,.
\end{equation}
The advective and parallel derivatives carry non-linearities entering
through $\phi$ and $\Psi$:
$ \vedl = \{\phi,\}, $ and $
  \dpl = \pss{} - \{\bhat\Psi,\}$.
Finally, the curvature operator ${K} = - \omega_{B} \left( \sin s \partial_x
        + \cos s \partial_y\right) $
originates from compressibility terms. 
The parameters in the equations reflect the competition between parallel
and perpendicular dynamics, reflected in the scale ratio
$\epss=(qR/L_\perp)^2$.  The electron parallel dynamics is controlled by
\begin{equation}
\label{eq:eqparms}
  \bhat={2 \mu_0 p_e\over B^2}\,\epss\,, \qquad\qquad
  \muhat={m_e\over M_i}\,\epss\,, \qquad\qquad
  C = 0.51 {L_\perp\over\tau_e c_s}\muhat = \nu \muhat\,,
\end{equation}
where $\tau_e$ is the electron collision time
and the factor  $0.51$ reflects the parallel
resistivity
\cite{Braginskii:1965}.  The competition between these three parameters,
representing magnetic induction, electron inertia, and
resistive relaxation determines the response of $J$ to the force imbalance in
Eq.~(\ref{eq:eqpsi}).
The simulations are performed on a grid with  
$64 \times 256 \times 16$ points and dimensions $64 \times 256 \times 2 \pi$ in ${x,y,s}$.
Some runs were repeated with double resolution  to ensure convergence. 
Standard
parameters for the runs were  $\hat \mu = 5$, $\hat s = 1$, $\omega_{B}
= 0.05  $ with the  viscosities set to
$\mu_w = \mu_n = 0.025$.
\\
A transition in the dynamical behaviour  from drift-Alfv\'en to turbulence dominated by MHD
modes is in this model triggered by increasing the ratio of $\hat \beta / \hat
\mu$.
To check the nature
of the observed turbulence  we evaluate the phase shift $\alpha$ 
between density and potential
fluctuations 
in dependence of $k_y$ using the relationship 
$\alpha_{k_y} = \log { I}(\phi_{k_y} n_{k_y}^*)$.  
For better statistics the phase is  averaged along the parallel
coordinate $s$ and time.
\begin{figure}[h!]
\begin{minipage}[t]{.475\textwidth}
\centering 
\includegraphics[width=\textwidth]{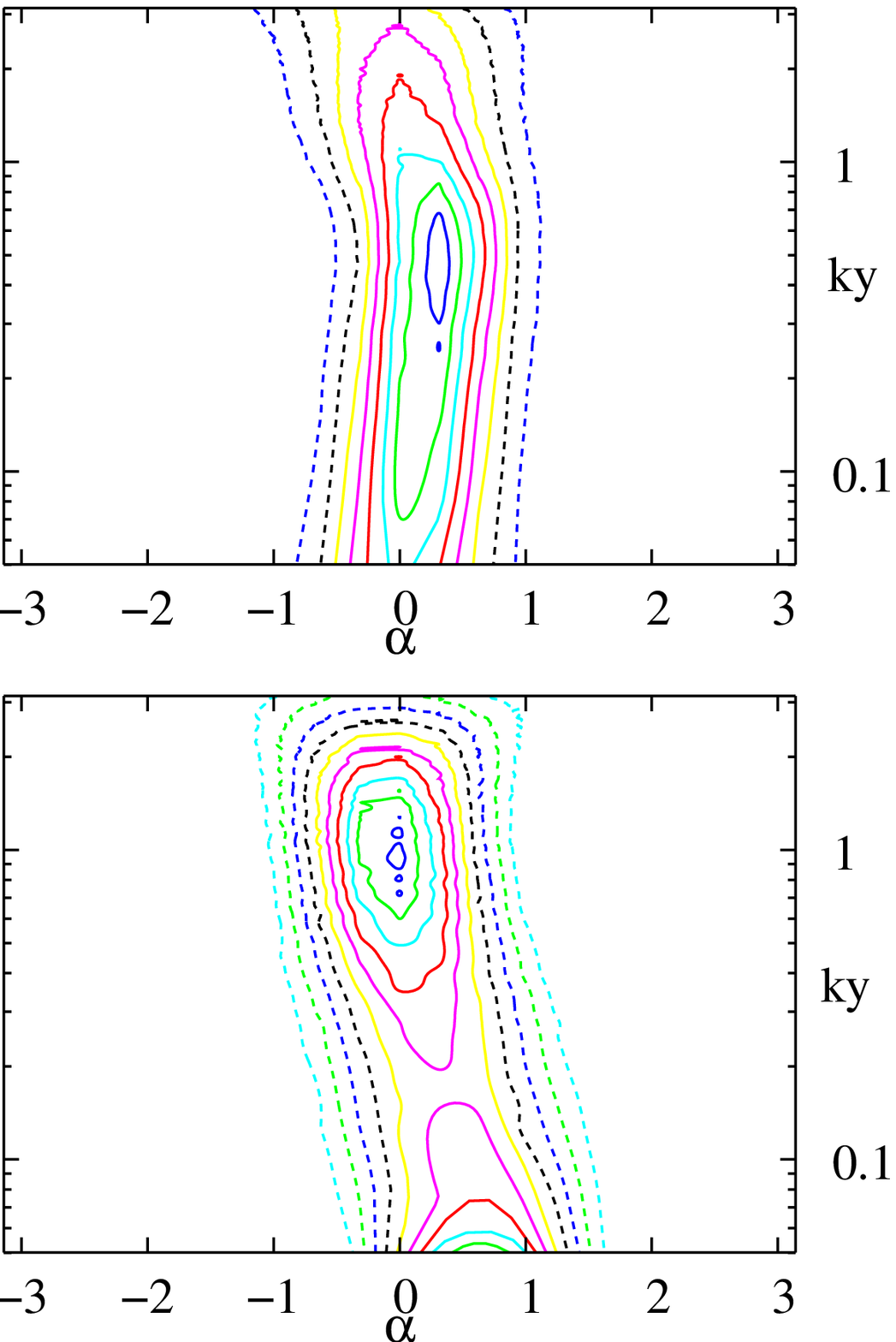}
\caption{Probability of phase angle $\alpha$ between density and potential fluctuations for
  $\hat \beta = 0.5$ (top) and $\hat \beta =50$ (bottom) versus $k_y$.
\label{Fig:PhaseAngle}}
\end{minipage}
\hfill
\begin{minipage}[t]{.475\textwidth}
\centering 
\includegraphics[width=1.1\textwidth]{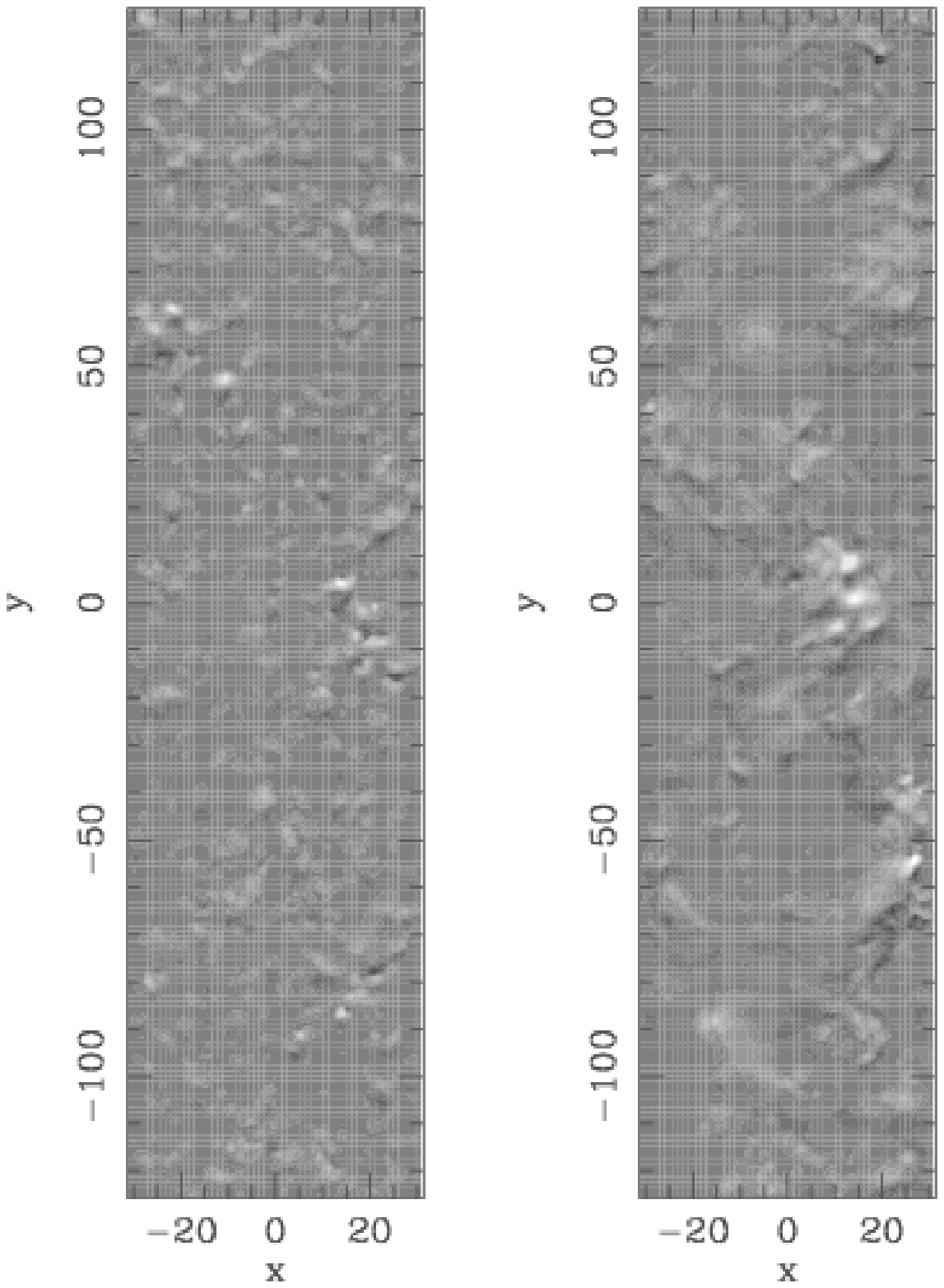}
\caption{Spatial structure of the transport $\Gamma_{n,\text{loc}}$  at the
  for $\hat \beta=0.5$ (left) and  $\hat
  \beta = 50$ (right). 
\label{FIG:transport_beta_space}}
\end{minipage}
\end{figure}
Fig.~\ref{Fig:PhaseAngle} shows for a low value 
 $\hat \beta = 0.5 $  and  a  high value $\hat \beta = 50$ 
the change in principal behaviour of the turbulence with increasing $\hat \beta$.
A shift towards larger phase angles is
observed  for low $k_y$-modes  
indicating the transition from drift-Alfv\'en to MHD type of turbulence.
%For the low beta case the phase shift between density and potential fluctuations is found
%to be small and slightly increasing with the poloidal wave number
%$k_y$ indicating the drift-Alfv\'en regime.  
As the turbulence character changes  to become more of the MHD type the
phase relationship between density and potential is altered. 
This is also reflected 
%in the $\hat \beta $ scaling of the $E \times B$
%flux show in Fig.~\ref{Fig:beta_scaling_flux}. 
%%
%\begin{figure}[h!]
%\centering 
%\includegraphics[width=0.75\textwidth]{turin6.2.eps}
%\caption{Scaling of $E \times B$ flux with $\hat \beta / \hat
%  \mu$. Increase of the flux with transition to the MHD regime is
%  observed. The bars indicate the width of the transport distribution
%  function.
%\label{Fig:beta_scaling_flux}}
%\end{figure}
%%
in a  visual inspection of the spatial transport structure (Fig.~\ref{FIG:transport_beta_space}).
It shows that the transport for high
values of $\hat \beta$ is carried by larger structures than for low
$\hat \beta$.
The local flux
$\Gamma_{n,\text{loc}}$ has the interesting structure that outward (white)
and inward (dark) transport regions are closely related, 
this reflects the fact that the transport is
carried in localized structures, having an outward flux at the one side
and an inward flux at the other side. They are linked to corresponding
eddies in the vorticity field. 
\\
In both regimes the turbulence decreases the equilibrium density gradient by about
$5-10\%$ in the stationary turbulence, a further deviation from
the background density gradient is prevented by a feed-back mechanism
in two damping layers at the inner and outer radial boundary, which
drive the flux-surface averaged density towards the initially specified
values. 
The  PDFs of transport related quantities for  $\hat \beta = 0.5$ and  $\hat \beta = 50$
are shown in Figs.~\ref{Fig:FluxPDF_lowbeta} and
\ref{Fig:FluxPDF_highbeta}.
In the low beta case density and potential fluctuations have PDFs that
are very well described by  Gaussians, with minor deviations from Gaussianity in
the tails.
Density and potential have a rather similar PDF as they are well
correlated and consequently
the fitted folded Gaussian $P_{FG}$  Eq.~(\ref{flux_G})  describes 
rather well the point-wise
flux PDF ($c=0.2$) . 
An indication of a power-law tail
in the point-wise flux PDF is not observed, and the point-wise flux
PDF decays exponentially. 
The 
PDF of the magnetic flux surface averaged density flux $\Gamma_{n,\text{FS}}$, shown in
Fig.~\ref{Fig:FluxPDF_lowbeta} c), is much
closer to a shifted Gaussian, depicted for comparison in the figure, but 
we observe  a  ``fat-tail''  towards larger transport events.
However, the flux surface averaged particle flux
will, even though not positive definite,  be at least  limited from
below, with excursions to a negative averaged flux being
unlikely.  
Thus in Fig.~\ref{Fig:FluxPDF_lowbeta} c) we also present the fitted
log-normal distribution, with parameters $\sigma = 0.273$, $z =-9.3$, and $m = 11$. 
 This fits the data well, but
over-predicts the tail,  an indication of the fact that the transport is limited
from above as well, as we are considering a finite
system. 
Indeed the EVD with parameters 
$h_0 = 2.60$ and $g=0.34$ 
fits the data very well, especially  the tail of the PDF. 
For the  MHD situation with  $\hat \beta = 50$  --- even though the
phase relation between density and potential differs and the turbulence
has different character  --- we
observe transport statistics similar to the drift-Alfv\'en case.
The corresponding PDFs for this case are depicted in
Fig.~\ref{Fig:FluxPDF_highbeta}. 
As the density and potential fluctuations are less well
coupled in that regime  the PDFs of  density and potential  differ
more.
The potential
fluctuations also deviate more from a Gaussian. Compared to the density
fluctuations they now have a larger width. Note that the fluctuation level has
increased significantly as well as the transport level when compared to
the low beta case. 
Nevertheless the PDF of the  point-wise measured
transport is still well described by a folded Gaussian as seen in
Fig.~\ref{Fig:FluxPDF_highbeta} b. The flux surface averaged transport is
shown in comparison with a fitted EVD and a Gaussian. 
The log-normal distribution function is not plotted here, as it is very close to the EVD. 
The flux surface averaged transport behaviour has changed
little as compared to the low beta case. It is still much closer to an
EVD (or log-normal) than to a Gaussian one, however, we observe
that to the low transport side the Gaussian distribution fits the
transport PDF better.
This effect is due to enhanced levels of small noisy  transport events
occurring in the MHD type of simulation and 
the flux PDF is more influenced by these  small scale ``random''
transport events. Their distribution centers around zero, while the
large scales reproduce the global transport PDF and carry the net
flux.  The small scale transport events then make an
appearance in the transport PDF   
for small averaged fluxes, which explains the nearly
Gaussian behaviour of the flux PDF on the low transport side.

\section{Global Models}
After having considered local models we now drop the scale separation
between turbulent fluctuations and equilibrium --- or in the cases where no
equilibrium exists, as in the SOL, between fluctuations
and time averaged background. A 
motivation for this is that  in the edge the fluctuation level
is of the same order of magnitude as the background and going out into
the SOL region the fluctuations in density can actually exceed the
average density by a large factor.  
However, regarding fluctuations  and background on the same footing
requires that the system is integrated to much longer times and thus
higher demands in terms of computational power. 
We thus consider a two dimensional and a simple three dimensional model.

\subsection{Flute Model}
Concerning only two dimensional dynamics of the plasma is -- if at all
-- only a good approximation in the SOL. We here consider a newly
developed global interchange model for the full particle density $n$,
temperature $T$ and vorticity $\vor$, with $\dtt = \pt_t + \frac{\hat z \times \nabla \phi \cdot \nabla}{B}$:
\begin{eqnarray}
\dtt{n} +  n {K}(\phi) -{K}(n) &=& \mu_n \nabla_\perp^2 n + S_n
- \sigma_n (n -1)
\\
\frac{3}{2} \dtt{ T} +  T  {K}(\phi)  - \frac{7}{2} T {K}(T) -
    \frac{T^2}{n}  {K}(n)  &=&  \mu_T \nabla_\perp^2 T + S_T
    - \frac{3}{2} \sigma_T (T-1)
\\
\left(\pt_t + \hat z \times \nabla \phi \cdot \nabla \right) \vor -
{K}(p)  &=& \mu_\vor \nabla_\perp^2 \vor -
\sigma_\vor \vor\;.
\end{eqnarray}
Here the coupling between the equations is exclusively due to the
curvature operator $K=\zeta\partial/\partial y$ where $\zeta=2\rho_s/R_0$
with $R_0$ the radius of curvature of the toroidal magnetic field. The
coupling due to parallel currents giving rise to drift-wave dynamics
is absent from these equations. 
\begin{figure}[h!]
\centering 
\includegraphics[width=0.7\textwidth]{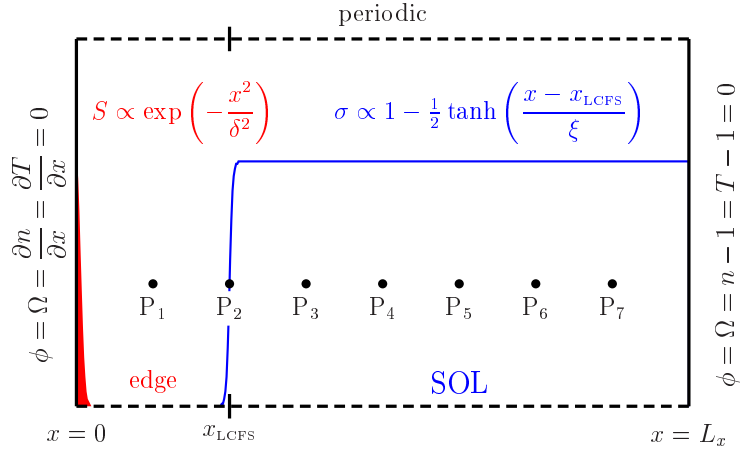}
\caption{
Geometry showing shape of source and sink terms and the locations were
long timeseries are taken.
\label{Fig:Geometry}}
\end{figure}
The $S$-terms on the right hand sides
represent sources of particles and heat, while the $\sigma$-terms
operating in the SOL (Fig.~\ref{Fig:Geometry}) are
sinks due to the parallel loss to end plates along open field lines.
Boundary conditions correspond to free-slip, while in
normalized units $n=T=1$ at $x=L_x$ and $\partial n/\partial x=
\partial T/\partial x=0$ at $x=0$, see Fig.\ref{Fig:Geometry}. For strong forcing the system is
unstable to interchange modes causing significant convective transport.
A novel property of this model is the non-linear conservation of the
energy integral
$
E = \int d\bm{x}\,\left[ \frac{1}{2} \left( \nabla\phi \right)^2 +
 \frac{3}{2}\,nT \right] 
$,
revealing the conservative transfer of energy from thermal to kinetic form
due to magnetic field curvature. A realistic modeling of this process is
mandatory for predictive global models. Nevertheless the solutions for
strong forcing shows the characteristic ``on-off'' character of the
turbulence due to self-sustained sheared flows \cite{Naulin:PRL:2003}.
The time-averaged thermal gradients are strongest in the source regions
and are ``flapping'' back and forth, leading to intermittent ejection
of particles and heat far out into the SOL region.
\begin{figure}[h!]
\centering 
\includegraphics[width=0.475\textwidth]{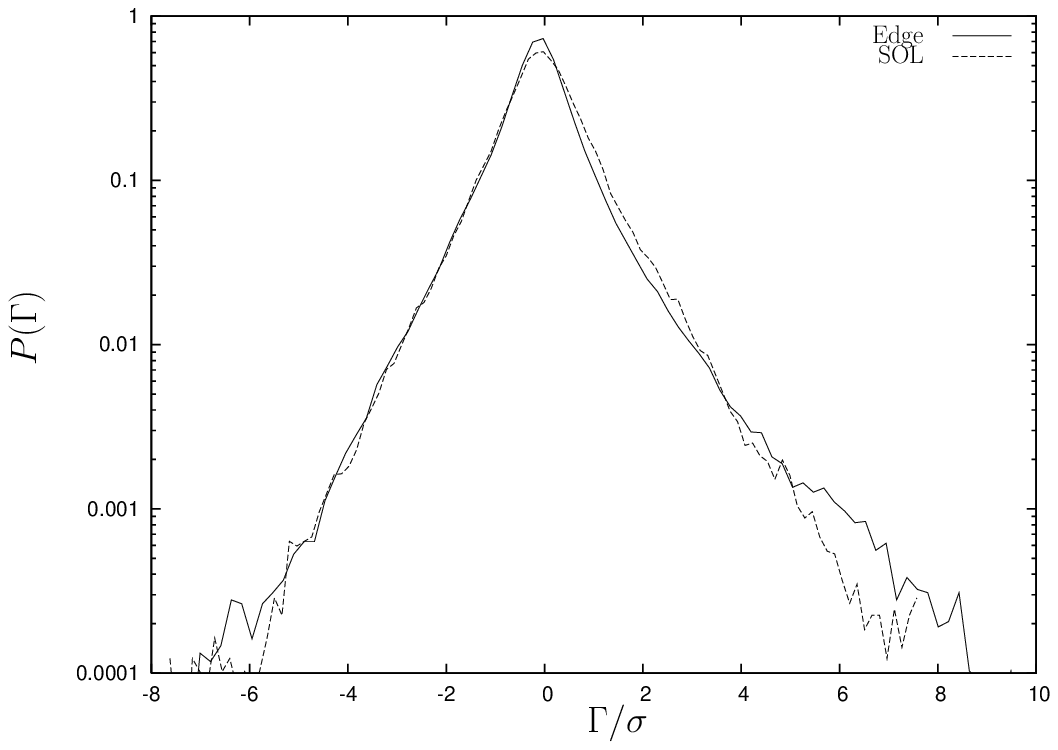}\hfill
\includegraphics[width=0.475\textwidth]{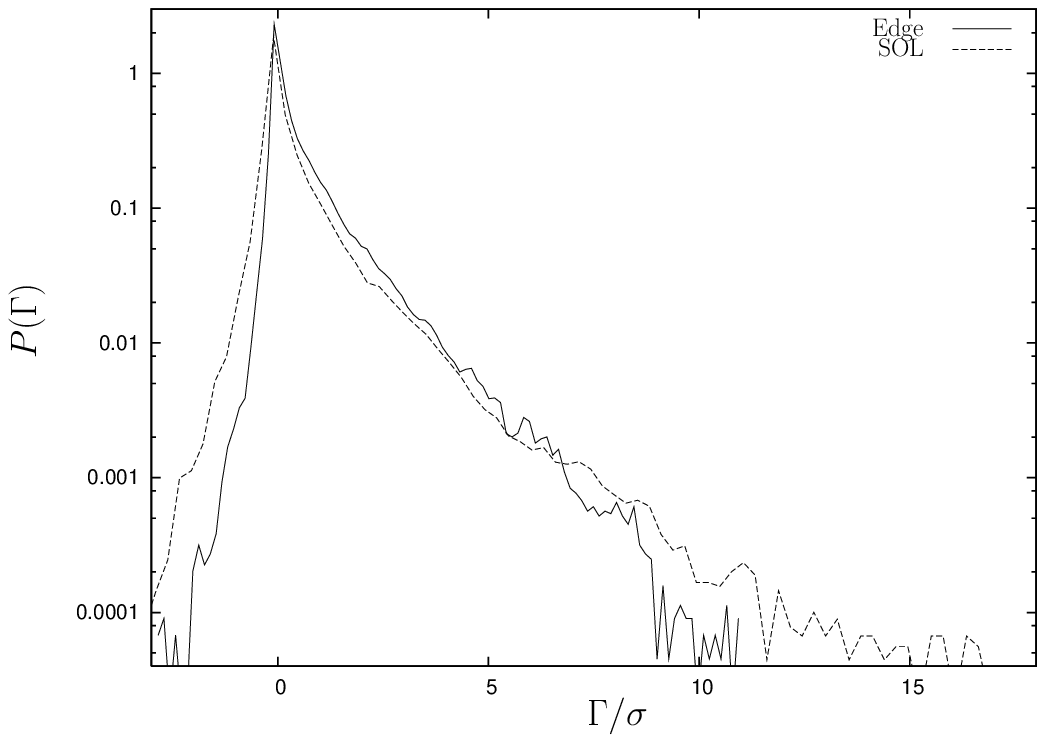}
\caption{
PDF of pointwise and flux surface averaged transport  inside LCFS and in the SOL
\label{Fig:FluxPDF_2DGlobal}}
\end{figure}
The transport PDF  obtained from direct numerical simulation of that
model is very different from the ones obtained for the
fluctuation based model. The flux surface averaged transport is no
longer constant and instead varies with radial position  as density is lost in the parallel
direction. 
We here take the transport data from a long timeseries since it is
impossible  to
use spatial homogeneity in the radial direction to increase the quality
of the statistics. 
It here makes  sense to compare flux PDF's by normalising
them to their root mean square. Fig.~\ref{Fig:FluxPDF_2DGlobal} shows
the PDFs of the point-wise and the flux-surface averaged transport at
radial positions corresponding to the edge and the SOL part of the
simulation domain, ($P_1$ and $P_4$ in Fig.~\ref{Fig:Geometry}).  A slight tendency to larger flux events in the SOL is observed. 
Averaging the transport poloidally has less dramatic effects than for
the models considered in Section 2. The
reason for this is the large poloidal 
correlation lengths of the order of the poloidal size of the simulation
domain.
\begin{figure}[h!]
\begin{minipage}[t]{.475\textwidth}
\centering 
\includegraphics[width=\textwidth]{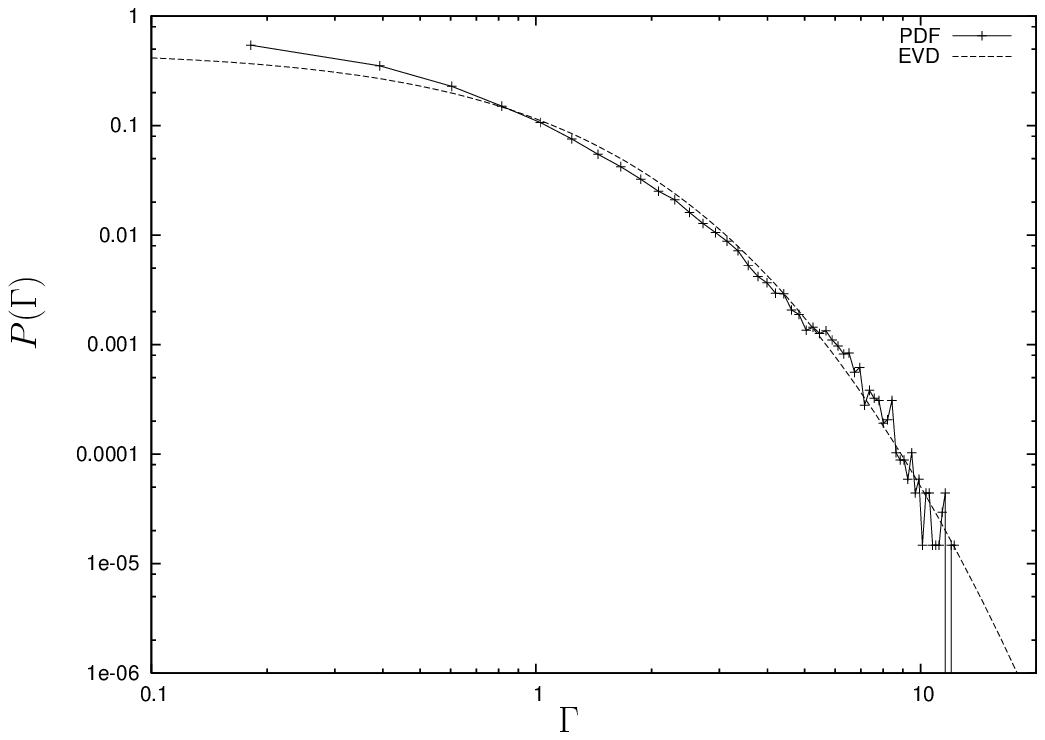}
\caption{
Tail of the flux surface averaged transport  in the SOL (see
Fig.~\ref{Fig:FluxPDF_2DGlobal}) fitted with an
EVD.
\label{Fig:FluxPDF_log_evd}}
\end{minipage}
\hfill
\begin{minipage}[t]{.475\textwidth}
\centering
\includegraphics[width=\textwidth]{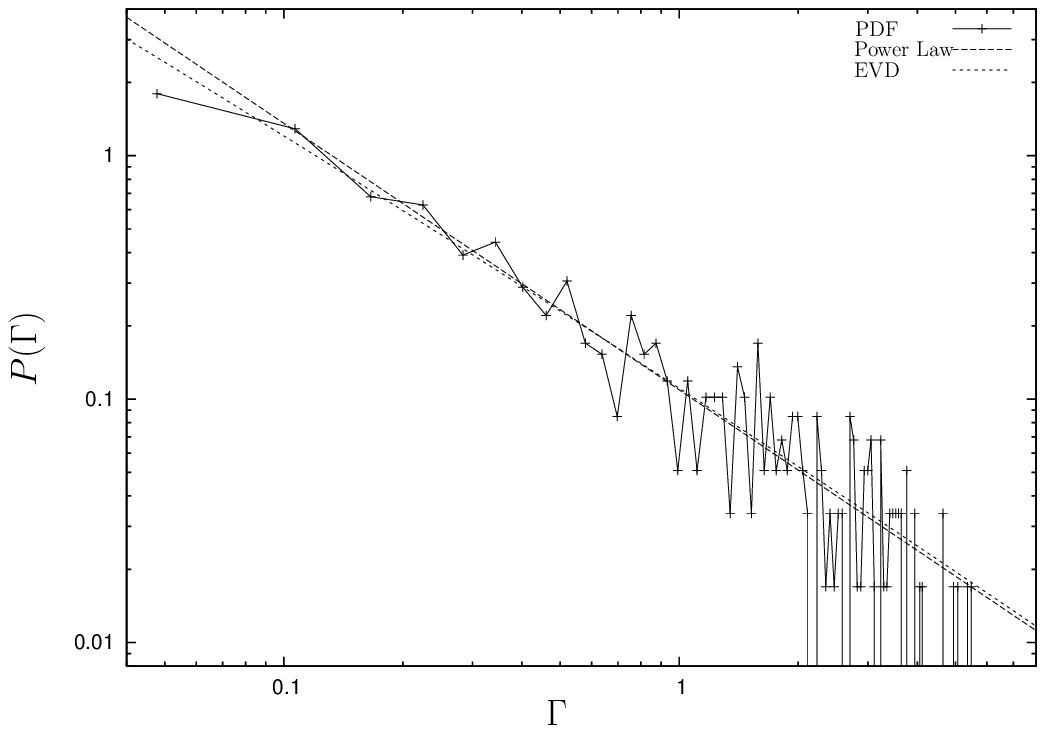}
\caption{
Flux surface and time averaged transport  in the SOL fitted with an
EVD and a power law.
\label{Fig:FluxPDF_log_pow}}
\end{minipage}
\end{figure}
Fig. \ref{Fig:FluxPDF_log_evd} demonstrates that even for
this obviously highly intermittent and bursty transport the EVD ($h_0
= 1.35$ and $g=0.60$ ) is an
excellent fit for the tail of the PDF. 
However, if the data are additionally
averaged over times longer than the times between bursts, we observe a
behaviour that can as well be approximated by a power law PDF as an
EVD (Fig.~\ref{Fig:FluxPDF_log_pow}). The amount of data being in the
tail is small in that case, and the data-range is about 1.5 decades in
$\Gamma$ and down from six to only two decades in the ordinate
(Fig.~\ref{Fig:FluxPDF_log_evd}).  
One should note that whenever the variance of the data is
large the body of a  EVD will be similar to a
straight line in a log-log plot, while its tail  decays 
with increasing slope in the log-log plot. This
shows the danger of fitting a PDF only for a very limited range in a
doubly logarithmic plot. 
Our interpretation is thus, that to observe a distinct behaviour of the tail
of the transport PDF one needs to be able to fit data over at least two
decades and that fitting data on smaller regions is highly arbitrary
and not suited to distinguish between different interpretations of the
data.
 
\subsection{Electrostatic Drift Model}
Finally we consider a three dimensional global version of drift-wave equations. 
The model  uses  quasi-neutrality and electron density
continuity, together with the parallel force balance equations for ions
and electrons. Parallel
convection is kept. No assumptions are made on the scale
length of the background gradient compared to the fluctuation scales,
e.g. the constraint
$\nabla n_0 \sim \nabla \tilde n \sim \epsilon \ll 1 $ as underlying
f.x.~the HWE Eqs.: (\ref{EQ:HWOM}) and (\ref{EQ:HWD}) is dropped.
The resulting system is written in terms of the logarithm of density
$N = \log(n)$: 
\begin{eqnarray}
\dtt{ \vor} + \nabla N \cdot \frac{d}{dt} \nabla \phi 
&=& 
% \mu_\vor \nabla N \cdot  \nabla \vor  + 
\mu_\vor  \nabla^{2} \vor  +   \nabla_{\|} (U - V) + (U - V) \nabla_{\|} N
\\
\dtt{ N} &=&   - \left( V \nabla_{\|} N + \nabla_{\|} V \right)
+ \mu_{n} \left( \nabla^{2} N + ( \nabla N)^{2}\right)
\\
\dtt{ U} &=& - \nabla_{\|} \phi - \nu_{e,i} U
\\
\dtt{ V} &=&   \frac{M}{m_e}  \nabla_{\|} \left( \phi - N \right) -
 \left(
  \nu_{ei} + \nu_{e,n} \right) V
\end{eqnarray}
The energetic coupling is  due to the
parallel current dynamics that is by the difference in parallel electron velocity $V$
and parallel ion velocity $U$. Now coupling due to curvature is absent
from the equations.
Again, the density is the full density evolving and is fed into the
system on the inner side, while on the radially outer third of the
system Bohm-sheath boundary conditions are implemented. The inner two
thirds of the radius have periodic boundary conditions in the parallel
direction. The $y$-direction is periodic and fixed values are imposed
on  field quantities at the outer radial boundary,
while the diffusive fluxes are set to zero at the inner radial
boundary by prescribing a zero radial gradient there. The system is
solved on a 128x64x30 grid. Due to the heavy computational load of
that system it could not be integrated to times as high as the 2D
system, which  influences the quality of the statistics. 
\begin{figure}[h!]
\centering
\includegraphics[width=0.475\textwidth]{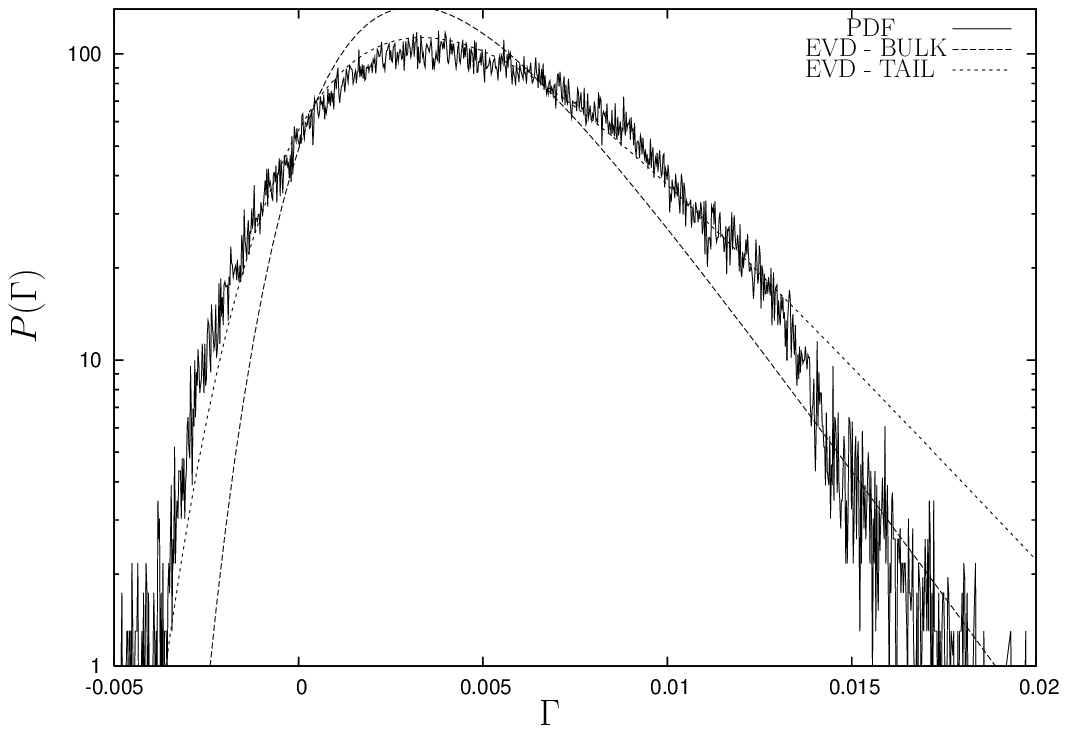}
\hfill
\includegraphics[width=0.475\textwidth]{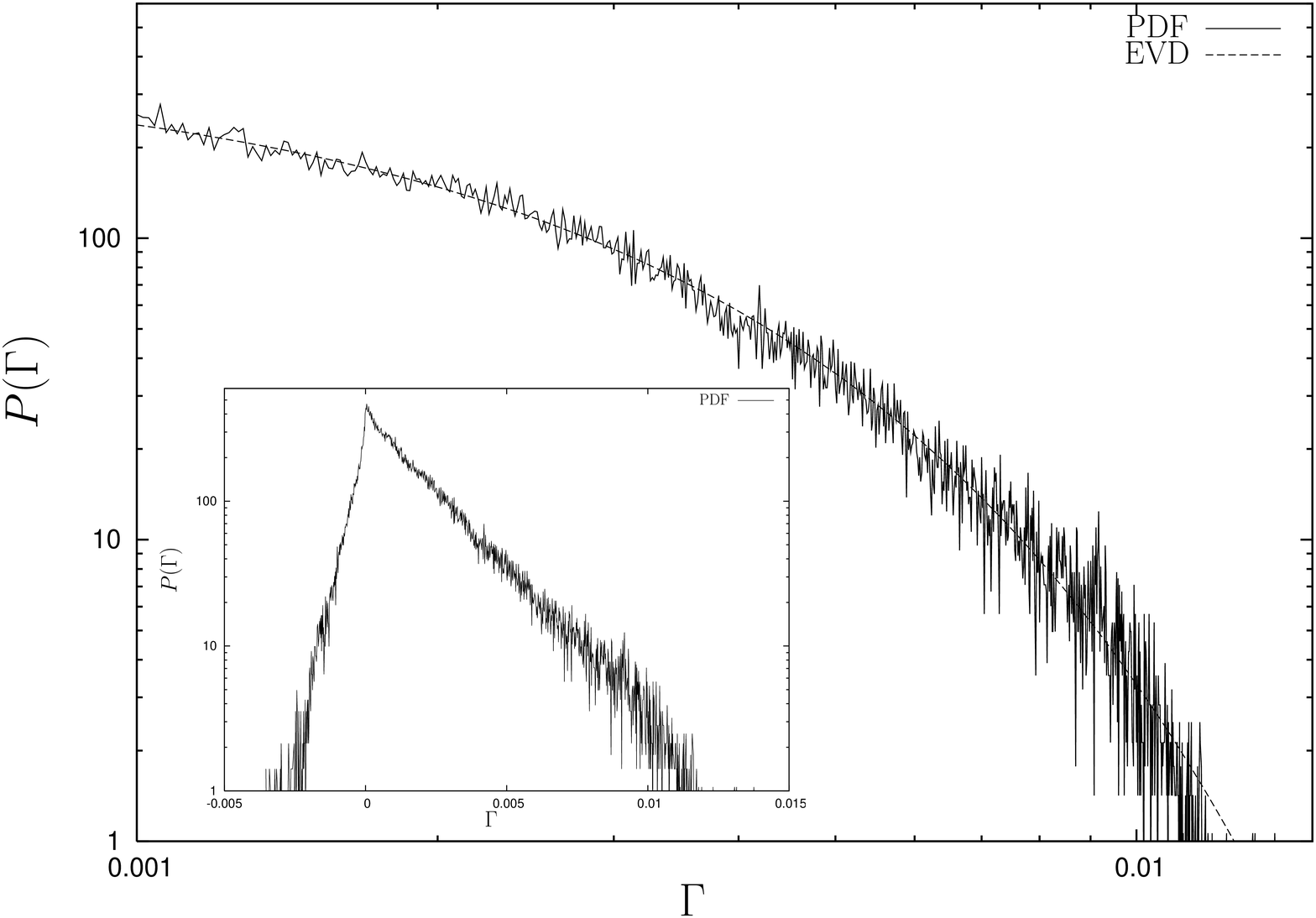}
\caption{
PDF of flux surface averaged transport in the edge (left) with fitted
EVD on bulk and tail of distribution 
 and in the SOL (right) with EVD fitted on the tail of the
 distribution.
\label{Fig:FluxPDF_HWGlobal}}
\end{figure}
The flux PDF in this case shows distinct differences between edge region
(closed field lines) and SOL  with open field lines and losses to the
limiter plates. For the edge region the poloidal correlation length is
shorter than the box length and thus the PDF of the averaged transport
is similar to that of the 2D- and 3D drift wave model. 
It fits well with an EVD in the
tail and in the bulk  (see
Fig.~\ref{Fig:FluxPDF_HWGlobal}). 
The flux PDF in the SOL part of the domain bears the characteristics
of  long poloidal 
correlations observed in that regime, due to the flute modes -- rather
than drift waves --
dominating  the dynamics in the SOL. 
Thus the shape of the transport PDF, also after flux surface averaging, is very
similar to that of the interchange model
(see. Figs.~\ref{Fig:FluxPDF_2DGlobal} and \ref{Fig:FluxPDF_log_evd}),
demonstrating also the applicability of 2D models in the SOL.
However, the tail of the PDF again does not
show any sign of a power law behaviour and can over the whole range be fitted very well
with an EVD.

\section{Conclusion}
Direct numerical simulations of physically distinct models of flute
and drift modes in two and three dimensional geometries have been
presented and analysed in terms of the transport PDF. 
For the fluctuation based models 
non-linear structures in the plasma dominate the transport, but 
give rise to a diffusive turbulent transport rather than a super-diffusion or
otherwise anomalous (in the fluid sense) transport. Consequently  a 
Gyro-Bohm like scaling of the transport with the magnetic field is
found. 
Similar behaviour has recently been reported
in investigations on test particle transport in plasma turbulence 
\cite{Basu:Naulin:Rasmussen:2003,Lin:Hahm:2002}.
\\
There is
little difference in the results concerning different physical models.
If the energy  input into the system is changed from a
homogeneous input due to a local, fixed pressure gradient to a
localized plasma source and sink regions are added the transport PDF 
becomes more bursty and
 longer poloidal
correlations arise.
Also in this situation the tails of all flux PDFs  considered are
extremely well fitted by extreme value distributions. 
Thus,
the deviation of the point-wise as well as the flux surface
averaged flux from Gaussianity is, at least in the asymptotic limit of
large system size, not due to anomalous --- in
the sense of non-diffusive --- behaviour. 
The presented  observations should not  be interpreted as indicators
for
the presence of self organized criticality.  An interpretation in
terms of  
occurrence of extreme transport events, which may be caused by localized  eddies
seems to be more appealing.
The statistics describing the transport PDF should therefor be taken
from extreme value statistics, which in all cases fits the tails of
the observed transport PDF very well.

\ack
This work was supported by the Danish Center for Scientific Computing
(DCSC), grants CPU-1101-08 and CPU-1002-17.
The authors  would like to thank B.D.~Scott for the introduction to
flux-tube geometries and for a number of detailed
discussions. 

\newpage 

\bibliographystyle{elsart-num}
\bibliography{/home/vona/TeX/pub/naulin,/home/vona/TeX/pub/drift,/home/vona/TeX/pub/dw,/home/vona/TeX/pub/numeric}

\end{document}